\documentstyle[12pt]{article}

 \font\tenmsb=msbm10 scaled\magstep 1
   \font\sevenmsb=msbm7 scaled \magstep 1
   \font\faivemsb=msbm5 scaled \magstep 1
\newfam\msbfam
      \textfont\msbfam=\tenmsb
      \scriptfont\msbfam=\sevenmsb
      \scriptscriptfont\msbfam=\faivemsb
\def\Bbb#1{{\fam\msbfam #1}}
\font\tengothic=eufm10 scaled\magstep 1
\font\sevengothic=eufm7 scaled\magstep 1
\newfam\gothicfam
      \textfont\gothicfam=\tengothic
      \scriptfont\gothicfam=\sevengothic

\newcommand{\bt}{\beta}
\newcommand{\lbd}{\lambda}

\newcommand{\om}{\omega}
\newcommand{\ra}{\rightarrow}
\newcommand{\prt}{\partial}
\newcommand{\be}{\begin{equation}}
\newcommand{\ee}{\end{equation}}
\newcommand{\dlt}{\delta}
\newcommand{\Dlt}{\Delta}
\newcommand{\al}{\alpha}
\newcommand{\ep}{\varepsilon}
\newcommand{\vp}{\varphi}
  
\newcommand{\gm}{\gamma}     

\begin{document}

\begin{center}
{\Large{\bf Self-Similar Structures and Fractal Transforms in 
Approximation Theory} \\ [5mm]

V.I. Yukalov$^1$ and E.P. Yukalova$^2$} \\ [3mm]

{\it $^1$ Bogolubov Laboratory of Theoretical Physics \\
Joint Institute for Nuclear Research, Dubna 141980, Russia \\

$^2$ Laboratory of Informational Technologies \\
Joint Institute for Nuclear Research, Dubna 141980, Russia}

\end{center}

\vskip 2cm

\begin{abstract}

An overview is given of the methods for treating complicated problems 
without small parameters, when the standard perturbation theory based 
on the existence of small parameters becomes useless. Such complicated 
problems are typical of quantum physics, many-body physics, physics of 
complex systems, and various aspects of applied physics and applied 
mathematics. A general approach for dealing with such problems has been 
developed, called {\it Self-Similar Approximation Theory}. A concise 
survey of the main ideas of this approach is presented, with the 
emphasis on the basic notion of group self-similarity. The techniques 
are illustrated by examples from quantum field theory. 

\end{abstract}

\newpage

\section{Preliminaries}

It would not be an exaggeration to say that practically all interesting 
realistic problems cannot be solved exactly, so that one almost always 
has to resort to some methods of approximate solution. In obtaining a 
solution, it is highly desirable, before plunging to numerical 
calculations, to get an approximate analytical solution to the problem, 
which could help to understand the basic properties and specific features 
of the considered case. It is just the possibility of deriving analytical 
presentations for approximate solutions that is our main concern in this 
paper. After deriving and studying such analytical presentations, nothing 
prohibits one to pass to a numerical procedure. Moreover, a prefactory 
analytical consideration can help in devising an approximate numerical 
algorithm, and the knowledge of the basic peculiarities of the problem 
under consideration can save plenty of computer time.

A general approach for treating complicated real world problems has 
been developed, called {\it Self-Similar Approximation Theory}. In this 
paper, we aim at delineating the principal ideas of the approach and 
in presenting its several new developments. To clearly distinguish the 
pivotal aspects of our theory from the characteristic points of other 
known techniques, we feel it necessary to say several words on the 
latter. There are, roughly speaking, three common ways of obtaining 
approximate solutions: single step estimates, asymptotic perturbation 
theory, and methods of successive iteration.

\vskip 3mm

{\bf A. Single Step Estimates}

\vskip 1mm

This kind of estimates is often related to minimizing or maximizing the 
corresponding part of an inequality. Probably, the most known and widely 
used such a tool is based on the Gibbs-Bogolubov inequalities, which are 
formulated as follows. Let $A$ and $B$ be Hermitian operators on a 
Hilbert space, for which the form
$$
F[A] \equiv - T\ln{\rm Tr} e^{-\bt A}
$$ 
exists, where $\bt T\equiv 1$, with $T$ being real. Define an average of 
$A$ with respect to $B$ as
$$
<A>_B \;  \equiv \frac{{\rm Tr}e^{-\bt B}A}{{\rm Tr}e^{-\bt B}} \; ,
$$
and similarly, an average of $B$ with respect to $A$. Then the 
{\it Gibbs-Bogolubov inequalities} are
\be
\label{1}
<A-B>_A \; \leq F[A] - F[B] \leq \; <A-B>_B \; .
\ee
This is the generalized presentation of the inequalities which are better 
known for the case when, instead of arbitrary Hermitian operators, one 
deals with Hamiltonians $H$ and $H_0$. In that case, the right-hand side 
of Eq. (1) becomes the inequality for the free energies,
$$
F[H] \leq F[H_0]\; +\; <H - H_0>_0 \; ,
$$
where $<\ldots>_0$ implies the averaging with respect to $H_0$. This inequality 
was derived by Gibbs [1] for classical statistics. Bogolubov [2] generalized 
it for quantum statistics and added the left-hand side of Eq. (1). When 
$T\ra 0$, the free energy reduces to the ground-state energy. Then, as a
particular case, one has the Peierls inequality. Introducing an effective
Hamiltonian
$$
H_{eff} \equiv H_0 \; + \; <H - H_0>_0 \; ,
$$
one may also write
$$
F[H] \leq F[H_{eff}] \; .
$$

In the standard way, one chooses the approximating Hamiltonian $H_0=H_0(\om_q)$
depending on a set of parameters or functions, say on a trial spectrum $\om_q$, 
so that $H_0$ could model the considered system and would allow one to calculate
the free energy $F[H_{eff}]$ for the effective Hamiltonian 
$H_{eff}=H_{eff}(\om_q)$. Then, one minimizes $F[H_{eff}]$ with respect 
to the trial functions $\om_q$ given by the equation
\be
\label{2}
\frac{\dlt}{\dlt\om_q}\; F[H_{eff}(\om_q)] = 0 \; .
\ee
When $<H-H_0>_0$ is positively defined, an approximate minimization can be 
done with the help of equation
\be
\label{3}
<H-H_0(\om_q)>_0 = 0 \; .
\ee

The Gibbs-Bogolubov inequality is constantly used in various problems of 
statistical mechanics and condensed matter theory. Among thousands of 
examples, let us mention the self-consistent photon approximation [3]. In 
quantum mechanics, the minimization of an effective ground-state energy 
with respect to trial parameters incorporated in a trial wave function is 
usually named the Ritz variational method. Such methods are 
termed single step since they give just a single approximation, without 
hinting on how to obtain subsequent corrections.

\vskip 3mm

{\bf B. Asymptotic Perturbation Theory}

\vskip 1mm

Contrary to the single step estimates, perturbation theory is a systematic 
procedure defining a sequence of approximations. It may, of course, happen 
for some complicated problems that one technically is able to calculate 
only a few terms of the perturbation sequence, but perturbation theory is 
systematic in the sense of prescribing a general way for calculating 
perturbative terms of arbitrary order. There is, especially in physical 
literature, quite a mess in the usage of the terms "perturbation theory" 
as opposed to "nonperturbative methods". Therefore, to avoid in what 
follows linguistic confusion, it is necessary to concretize several 
principal points and definitions.

The {\it standard perturbation theory} presupposes the existence of 
small parameters permitting one to present solutions in the form of 
asymptotic series [4]. Because of the latter, the standard perturbation 
theory may be named {\it asymptotic perturbation theory}. Usually, when 
talking about perturbation theory, one keeps in mind exactly the standard 
asymptotic perturbation theory.

One can distinguish three main types of asymptotic series occurring 
in perturbation theory. Suppose the problem is in calculating a real 
function $f(x)$ of a variable $x\in\Bbb{X}\subset\Bbb{R}$. The case of one 
function of one variable is taken just for the simplicity of notations 
and is not principal. In general, the function $f(x)$ can depend on 
any number of other variables, but we separate and explicitly write down 
only one variable that is assumed to play the role of a small parameter.
Perturbation theory with respect to a small parameter $|x|\ll 1$ yileds 
a sequence of approximants $\vp_k(x)$, where $k=0,1,2,\ldots$, 
approximating the sought function $f(x)$ in the vicinity of $x=0$,
\be
\label{4}
f(x) \simeq \vp_k(x) \qquad (|x|\ll 1)\; .
\ee
The approximants $\vp_k(x)$ have the structure of asymptotic series of 
one of the following types.

\vskip 1mm

(i) {\it Expansion over small parameters}:
\be
\label{5}
\vp_k(x) =\sum_{n=0}^k a_n x^n \; .
\ee
Generally, this can be an expansion over one or several parameters.

\vskip 1mm

(ii) {\it Expansion over asymptotic sequences}:
\be
\label{6}
\vp_k(x) = \sum_{n=0}^k a_n \ep_n(x) \; .
\ee
Here, the sequence $\{\ep_n(x)\}$ is asymptotic in the sense of Poincar\'e, 
so that
\be
\label{7}
\left | \frac{\ep_{n+1}(x)}{\ep_n(x)}\right | \ra 0 \qquad (x\ra 0) \; .
\ee
In particular, $\ep_n(x)$ can be $\ep^n(x)$, with $\ep(x)$ being a given 
function of $x$.

\vskip 1mm

(iii) {\it Generalized asymptotic expansion}:
\be
\label{8}
\vp_k(x) =\sum_{n=0}^k a_n(x)\ep_n(x) \; .
\ee
In the latter, the coefficients $a_n(x)$ retain dependence on $x$ in order 
to satisfy some additional conditions, but so that the sequence 
$\{ a_n(x)\ep_n(x)\}$ be asymptotic.

The general feature of all the cases above is that the difference
\be
\label{9}
\Dlt\vp_k(x) \equiv \vp_k(x) -\vp_{k-1}(x)
\ee
forms an asymptotic sequence $\{\Dlt\vp_k(x)\}$, such that
$$
\left | \frac{\Dlt\vp_{k+1}(x)}{\Dlt\vp_k(x)}\right | 
\ra 0 \qquad (x\ra 0) \; .
$$
The construction of generalized asymptotic expansions can often be rather 
elaborate, aiming at improving the accuracy of calculations. For example, 
in the Lindstedt-Poincar\'e method [5], one expands over a small parameter 
the sought solution and the frequency choosing the expansion coefficients 
so that to kill secular terms. In the Krylov-Bogolubov averaging technique 
[6], the generalized asymptotic expansion is constructed above an initial 
approximation including nonlinearity in order to model well the main 
properties of a nonlinear system. In the theory of anharmonic crystals [7,8], 
the expansion in powers of a small anharmonic parameter starts with a 
self-consistent phonon approximation partly taking account of anharmonicity. 
Nevertheless, no matter how elaborate is an initial approximation and how 
complicated is the structure of the resulting generalized expansion, all 
abovementioned cases presuppose the existence of small parameters and are 
typical examples of asymptotic perturbation theory.

\vskip 3mm

{\bf C. Methods of Successive Iteration}

\vskip 1mm

For numerical iterative algorithms, the existence of small parameters is not 
required. When the considered equation can be presented in the form
$$
Af(x) = h(x) \; ,
$$
in which $A$ is an operator and $h(x)$, a given function, then an iterative 
solution of this equation reads
\be
\label{10}
\vp_{k+1}(x) =\vp_k(x) + B_k\left [ A\vp_k(x) - h(x)\right ] \; ,
\ee
where $B_k$ are operators chosen so that to make the sequence $\{\vp_k(x)\}$ 
convergent [9]. The iterative algorithm (10) leads to approximants 
$\vp_k(x)$ that, in general, are not asymptotic series, even if $x\ra 0$. 
This is because an expansion of $\vp_k(x)$ at $x=0$, provided it exists, 
has the structure of a series 
$$
\vp_k(x) = \sum_{n=0}^k a_n^k \; x^n \; ,
$$
with coefficients $a_n^k$ labelled by two indices. Then the difference (9) 
is
$$
\Dlt\vp_k(x) = \sum_{n=0}^{k-1}\left ( a_n^k - a_n^{k-1}\right ) x^ n +
a_n^k \; x^k \; ,
$$
which is not of order $x^k$ but contains lower powers of $x$, since 
$a_n^k\neq a_n^{k-1}$. Hence, $\{\Dlt\vp_k(x)\}$ is not an asymptotic 
sequence.

Though the method of successive iteration looks more general than 
perturbation theory, it has a weak point of being, in the majority of cases, 
purely numerical, without providing analytical formulas. And also, for very 
complicated problems a numerical procedure can be too much time consuming 
or even unsolvable.

\section{Optimized Perturbation Theory}

A qualitatively different systematic approach was advanced [10] for 
treating the problems without small parameters and permitting one to obtain, 
at least in first orders, approximate analytical solutions. This approach was 
motivated by and applied to many-particle systems with strong interactions 
[11--21]. Technically, the approach is based on the methods of perturbation 
and iteration theories combined with optimal control theory.

The pivotal idea of the approach [10] is the introduction of {\it control 
functions} whose role is to govern the convergence of approximation 
sequences. Generally, an approximation theory consists of three parts: 
an initial approximation, a calculational algorithm, and a sequence 
analysis. Control functions can be introduced in any of these parts. The 
main is that the resulting approximation sequence $\{ F_k(x,u_k)\}$ be 
convergent due to control functions $u_k=u_k(x)$. {\it Optimized approximants} 
are 
\be
\label{11}
f_k(x) \equiv F_k(x,u_k(x)) \qquad (k=0,1,2,\ldots)\; ,
\ee
forming a convergent sequence $\{ f_k(x)\}$. In obtaining the approximations 
$\{ F_k(x,u_k)\}$, it is not required to have any small parameters, but, if 
the techniques of perturbation theory are employed, an expansion is made 
with respect to a formal parameter $\ep$ that at the end is set to unity
$$
F_k(x,u_k) = \sum_{n=0}^k a_{nk}(x)\ep^n \qquad (\ep\ra 1) \; .
$$
This expansion is not asymptotic, even when $x\ra 0$, since the coefficients 
$a_{nk}$ are labelled by two indices. In this way, the expressions $F_k(x,u_k)$,
although can be obtained by means of formal techniques of perturbation theory, 
have the structure typical of the terms of an iterative procedure. Thus, the 
approach is not an asymptotic perturbation theory. Moreover, the approximations
$F_k(x,u_k)$ may also be obtained by means of an iterative algorithm. 
Therefore, the optimized perturbation theory is principally different from 
the standard asymptotic perturbation theory, since the former does not require 
any small parameters and the structure of its approximate terms is not that of 
asymptotic series.

{\bf Definition}. {\it Optimized Perturbation Theory} is a systematic method 
defining a sequence of successive approximants, whose convergence is governed 
by control functions.

How could we formulate general rules for finding such control functions? Since 
the role of the latter is to govern convergence, let us write down the Cauchy 
criterion of uniform convergence for the sequence $\{ F_k(x,u_k)\}$, with 
$x\in\Bbb{X}$. The sequence is uniformly convergent on $\Bbb{X}$ if and only 
if for each given $\ep$ there exists $N_\ep$ such that
\be
\label{12}
|F_{k+p}(x,u_{k+p}) - F_k(x,u_k)| < \ep
\ee
for all $k\geq N_\ep$ and $p\geq 1$. We also wish that convergence be as 
fast as possible. A general way of defining control functions, providing for 
the considered system the required property, is formulated by the {\it optimal 
control theory} [22]. According to this theory, control functions are given 
by minimizing a cost functional. In our case, the cost functional garanteeing 
the fastest convergence can be constructed as the {\it fastest-convergence 
cost functional}
\be
\label{13}
{\cal F}_u =\frac{1}{2}\sum_{n=0}^\infty \left [ F_{n+p}(x,u_{n+p}) -
F_n(x,u_n)\right ]^2 \; ,
\ee
in line with the Cauchy criterion (12). The fastest convergence is achieved 
by means of control functions, minimizing the cost functional (13), that is, 
from the condition 
\be
\label{14}
\frac{\dlt{\cal F}_u}{\dlt u_k} = 0 \; , \qquad 
\frac{\dlt^2{\cal F}_u}{\dlt u_k^2} > 0\; .
\ee
The variational derivatives are
$$
\frac{\dlt{\cal F}_u}{\dlt u_k} = (2F_k - F_{k+p} - F_{k-p})F_k' \; ,
$$
\be
\label{15}
\frac{\dlt^2{\cal F}_u}{\dlt u_k^2} = 2(F_k')^2 + (2F_k - F_{k+p} - 
F_{k-p})F_k'' \; ,
\ee
where $F_k=F_k(x,u_k)$, $1\leq p\leq k$, and
$$
F_k' \equiv \frac{\prt F_k}{\prt u_k} \; , \qquad
F_k'' \equiv \frac{\prt^2 F_k}{\prt u_k^2} \; .
$$
The first of Eqs. (14) possesses two solutions yielding two possible 
optimization conditions.

\vskip 1mm

(i) {\it Differential optimization condition}
\be
\label{16}
\frac{\prt}{\prt u_k} \; F_k(x,u_k) = 0
\ee
gives an extremum of the cost functional, but it is not clear what this extremum
is, since the sign of the second derivative
$$
\frac{\dlt^2{\cal F}_u}{\dlt u_k^2} =  (2F_k - F_{k+p} - F_{k-p})F_k''
$$
is not defined. To concretize the situation, we need to invoke some additional
constraints. Consider a particular case of $p=1$ and assume the validity of the
{\it fast convergence condition}
\be
\label{17}
F_k(x,u_k) \approx F_{k+1}(x,u_{k+1}) \; .
\ee
Then the differential condition (16) is an approximate condition for the minimum
of the cost functional, since
$$
\frac{\dlt^2{\cal F}_u}{\dlt u_k^2} \approx 0 \; , \qquad 
\frac{\dlt^3{\cal F}_u}{\dlt u_k^3} \approx 0 \; , \qquad
\frac{\dlt^4{\cal F}_u}{\dlt u_k^4} \approx 6(F_k'')^2 > 0 \; .
$$

\vskip 1mm

(ii) {\it Difference optimization condition}
\be
\label{18} 
 F_{k+p} - 2F_k + F_{k-p} = 0
\ee
clearly correspond to the minimum of the cost functional, as far as
$$
\frac{\dlt^2{\cal F}_u}{\dlt u_k^2} = 2(F_k')^2 > 0 \; .
$$
However, this does not uniquely define control functions, since Eq. (18) 
contains three of them, $u_{k+p}, \; u_k$, and $u_{k-p}$. To resolve the problem, 
consider again the case of $p=1$ and assume that the value $F_k(x,u_k)$ weakly 
depends on the change of $u_k$ by $u_{k+1}$, which can be formulated as the {\it 
weak sensitivity condition}
\be
\label{19}
F_k(x,u_k) \approx F_k(x,u_{k+1}) \; .
\ee
Then the difference condition (18) reduces to the equation
\be
\label{20}
F_k(x,u_k) - F_{k-1}(x,u_k) = 0 \; ,
\ee
making it possible to define $u_k=u_k(x)$.

In this way, neither Eq. (16) nor Eq. (18) can serve as exact equations 
for minimizing the cost functional (13) and for uniquely defining control 
functions. But the latter are unambiguously defined and the cost functional 
(13) is approximately minimized by either optimization condition (16) or 
condition (20). These conditions are equivalent to each other. Both of them 
are approximate. Both are variational, following from the variation of a 
cost functional. Both invoke the notion of weak sensitivity with respect to 
the variation of control functions. Both are designed for supplying fastest 
convergence of the sequence of optimized approximants by minimizing the 
fastest-convergence cost functional. Being completely equivalent by their 
meaning, the optimization conditions (16) and (20) differ only by their 
form, being expressed either through a derivative or through
 a finite difference.

Solving one of the optimization equations, either (16) or (20), one gets 
control functions $u_k=u_k(x)$ governing convergence of the optimized 
approximants (11). It may happen for some $k$, that an optimization equation 
does not have an exact solution. Then, since the optimization conditions 
themselves are approximate, it is admissible to look for an approximate 
solution for a control function. For example, if the latter becomes complex, 
one may take only its real part. Or, when neither the derivative (16) nor 
difference (20) are exactly zero, one may look for control functions 
minimizing one of the following forms:
$$
\min_{u_k}\left | \frac{\prt}{\prt u_k}\; F_k(x,u_k)\right | \; , \qquad
\min_{u_k}\left | F_k(x,u_k) - F_{k-1}(x,u_k) \right | \; .
$$

As is mentioned at the beginning of this section, control functions can 
be incorporated at any step of the theory. A straightforward way is to 
introduce them in the initial approximation. For instance, if one considers 
a problem described by a Hamiltonian $H$, one may take for the initial 
approximation a Hamiltonian $H_0(u)$ containing a set of trial parameters $u$. 
Then the Hamiltonian of the problem can be presented as
$$
H = H_0(u) + \ep [ H - H_0(u)] \; ,
$$
with a formal parameter $\ep\ra 1$. Accomplishing perturbative calculations 
with respect to the formal parameters $\ep$, one sets it to unity and obtains 
perturbative terms $F_k(x,u)$. After this, one finds control functions from 
an optimization condition and substitute them into $F_k(x,u_k)$. Absolutely
the same procedure can be realized if the considered problem is characterized 
not by a Hamiltonian but by a Lagrangian or an action. It is also possible 
to incorporate trial parameters into an initial approximation for a wave 
function or for Green functions and to derive the subsequent approximations 
by means of an iterative procedure.

As is evident, there can be a variety of particular technical ways of 
introducing initial approximations and control functions. But all such 
variants are based on the same fundamental idea of control functions 
governing convergence for the sequence of optimized approximants [10]. Several 
years after Ref. [10], there appeared a series of papers [23--28] advertizing 
the same idea of introducing control functions for rendering perturbation 
theory convergent. Nowdays the optimized perturbation theory is widely used 
for various problems, being employed under different guises and called by 
different names, such as modified perturbation theory, renormalized 
perturbation theory, variational perturbation theory, controlled perturbation 
theory, self-consistent perturbation theory, oscillator-representation method, 
delta expansion, optimized expansion, nonperturbative expansion, and so on 
[23--32]. Many problems of quantum mechanics, statistical mechanics,
condensed matter physics, and quantum field theory are successively treated 
by this optimized approach. Not to list numerous applications, let us cite 
a couple of recent reviews [33,34].

Despite a number of very successful applications, optimized perturbation 
theory does not provide answers to the following important questions:

\vskip 1mm

(i) How to improve the accuracy with a given number of approximate terms?

\vskip 1mm

(ii) How to realize a step-by-step control of the stability of the method, 
if no exact solutions for the problem are known?

\vskip 1mm

(iii) How to choose the best initial approximation, if several of them are 
admissible?

\vskip 1mm

To answer these questions it is necessary to look at the problem from the 
point of view of a more general approach, which is described in the next 
section.

\section{Self-Similar Approximation Theory}

An approach, more general than optimized perturbation theory, has been 
developed [35--57], being based on the techniques of the theory of dynamical 
systems and optimal control theory. The underlying idea of the approach is 
to consider the passage from one successive approximation to another as the 
motion on the manifold of approximations. Then, the approximation order 
$k=0,1,2,\ldots$ should play the role of discrete time, and each approximant 
should represent a point of a trajectory in the phase space of approximants.

To correctly describe evolution, one has to construct a dynamical system. 
For this purpose, we again need to introduce control functions. Employing 
some variant of the method of successive approximations, we get the terms 
$F_k(x,u_k)$. The role of control functions $u_k=u_k(x)$ is to make 
convergent the sequence $\{ f_k(x)\}$ of the optimized approximants 
\be
\label{21}
f_k(x) \equiv F_k(x,u_k(x)) \; .
\ee
Convergence implies the existence of the limit
\be
\label{22}
\lim_{k\ra\infty} f_k(x) = f^*(x) \; .
\ee
Let all approximating functions $f_k(x)$, with $k=0,1,2,\ldots$ and 
$x\in\Bbb{X}$, together with the limit (22) pertain to a complete space 
${\cal A}$, which we call {\it approximation space} and which plays the 
role of a {\it phase space} for the evolution of $f_k(x)$ with 
respect to the discrete time $k$.

Introduce a function $x_k(\vp)$ by the {\it reonomic constraint}
\be
\label{23}
F_0(x,u_k(x)) =\vp \; , \qquad x=x_k(\vp) \; .
\ee
Changing the variables, we define
\be
\label{24}
y_k(\vp) \equiv f_k(x_k(\vp)) \; .
\ee
The transformation $y_k: \; {\cal A}\ra{\cal A}$ is an endomorphism of 
the phase space ${\cal A}$, with a unitary element given by the equation
\be
\label{25}
y_0(\vp)=\vp \; .
\ee
By definition (24), each $f_k(x)$ corresponds to $y_k(\vp)$ and, conversely, 
to each $y_k(\vp)$ we set in correspondence
\be
\label{26}
f_k(x) = y_k(F_0(x,u_k(x))) \; .
\ee
By this construction, the sequences $\{ y_k(\vp)\}$ and $\{ f_k(x)\}$ are 
bijective. The existence of the limit (22) implies the existence of the 
limit
\be
\label{27}
\lim_{k\ra\infty} y_k(\vp) = y^*(\vp) \; ,
\ee
so that 
\be
\label{28}
f^*(x) = y^*(F_0(x,u^*(x)) \; ,
\ee
where $u^*(x)=\lim_{k\ra\infty}u_k(x)$. For the map $\{ y_k(\vp)\}$, the 
limit $y^*(\vp)$ is a fixed point, when
\be
\label{29}
y_k(y^*(\vp))=y^*(\vp) \; .
\ee
The dependence of the fixed point $y^*(\vp)$ on the starting point $\vp$ is 
due to the reonomic constraint (23).

A particular form of the endomorphism (24) depends on the choice of an 
initial approximation $F_0(x,u)$ and of control functions $u_k(x)$. These 
are to be chosen so that to guarantee the fastest convergence of the sequence 
$\{ y_k(\vp)\}$. Uniform convergence on ${\cal A}$ implies the validity of 
the Cauchy criterion 
\be
\label{30}
|y_{k+p}(\vp) - y_k(\vp) | < \ep
\ee
for all $k\geq N_\ep$ and $p\geq 1$. This suggests that the evolution in 
the map $\{ y_k(\vp)\}$ is to be such that to minimize the fastest-convergence 
cost functional
\be
\label{31}
{\cal F}_y =\frac{1}{2}\sum_{n=0}^\infty \left [ y_{n+p}(\vp) -
y_n(\vp)\right ]^2 \; .
\ee
The minimization is to be done with respect to any $y_n(\vp)$, including 
$y_0(\vp)=\vp$. As is evident, the absolute minimum of the functional (31) 
is realized if and only if the initial point $\vp$ coincides with the fixed 
point $y^*(\vp)$, when
\be
\label{32}
{\cal F}_y= 0\; , \quad \vp = y^*(\vp) \; .
\ee

{\bf Proposition}. For the fastest-convergence cost functional to be minimal 
it is necessary that the {\it self-similarity relation}
\be
\label{33}
y_{k+p}(\vp) = y_k(y_p(\vp))
\ee
be valid.

\vskip 1mm

{\bf Proof}. The absolute minimum of the cost functional (31) is ${\cal F}_y=0$. 
This is realized if and only if $\vp=y^*(\vp)$, when equation (33) becomes an
identity.

Clearly, the self-similarity relation (33) is a necessary but not sufficient 
condition for minimizing the cost functional (31). This relation describes 
the property of self-similarity in the general sense, which includes as a 
particular case the scaling self-similarity $y_k(\lbd\vp)=\lbd^{\al_k}y_k(\vp)$ 
taking place for the power-law functions $y_k(\vp)=\vp^{\al_k}$, with the powers 
such that $\al_{k+p}=\al_k\cdot\al_p$. The scaling self-similarity is what one 
usually keeps in mind when mentioning this property. However, this type of 
self-similarity is just a particular trivial case of the relation (33). The 
latter for the endomorphism $y_k$, together with the unitary element (25), 
defines the semigroup properties
\be
\label{34}
y_{k+p} = y_k\cdot y_p \; , \qquad y_0 = 1\; ,
\ee
because of which the relation (33) can be called the {\it group self-similarity}. 
In the theory of dynamical systems [58,59], the family of endomorphisms $\{ y_k|\;
k\in\Bbb{Z}_+\}$, where $\Bbb{Z}_+\equiv\{ 0,1,2,\ldots\}$, is termed a cascade,
which is a dynamical system in discrete time. Since in our case this family of
endomorphisms is formed of the sequence $\{ y_k(\vp)\}$ of the approximants 
$y_k(\vp)$, the corresponding cascade can be named the {\it approximation 
cascade}.

To deal with discrete time is less convenient than with continuous time, when the
latter is given on $\Bbb{R}_+\equiv [0,\infty)$. Therefore, it is useful to embed 
the cascade into a flow. The {\it embedding}
\be
\label{35}
\{ y_k|\; k\in\Bbb{Z}_+\} \subset \{ y(t,\ldots)| \; t\in\Bbb{R}_+\}
\ee
implies that the flow possesses the same semigroup property
\be
\label{36}
y(t+t',\vp) = y(t,y(t',\vp))
\ee
and the trajectory $\{ y(t,\vp)\}$ of the flow passes trough all points of the 
cascade,
\be
\label{37}
y(t,\vp) = y_k(\vp) \qquad (t=k\in\Bbb{Z}_+) \; .
\ee
The flow embedding the approximation cascade is termed the {\it approximation 
flow}.

The advantage of dealing with a flow is that differentiating the self-similarity
relation (36) one comes to a differential Lie equation
\be
\label{38}
\frac{\prt}{\prt t}\; y(t,\vp) = v(y(t,\vp)) \; ,
\ee
where $v(y)$ is a velocity field. Integrating the evolution equation (38) 
between $y_{k-1}=y_{k-1}(\vp)$ and $y_k^*=y_k^*(\vp)$, where $y^*_k$ is an 
approximate fixed point, we get the {\it evolution integral}
\be
\label{39}
\int_{y_{k-1}}^{y_k^*} \frac{dy}{v(y)} =\tau_k \; ,
\ee
with $\tau_k$ being the {\it effective approximation time} required for reaching 
the quasifixed point $y_k^*$. For short, we may call $\tau_k$ the {\it control 
time}. Due to the relations (23) to (28), the integral (39) may be presented as
\be
\label{40}
\int_{f_{k-1}}^{f_k^*} \frac{d\vp}{v_k(\vp)} =\tau_k \; ,
\ee
where $f_k=f_k(x)$, $f_k^*=f_k^*(x)$, with 
\be
\label{41}
f_k^*(x) \equiv y_k^*(F_0(x,u_k(x)) \; ,
\ee
and $v_k(\vp)$ is the cascade velocity given by a discretization of the flow 
velocity. The Euler discretization of the flow velocity is defined as
\be
\label{42}
v_k(\vp) \equiv V_k(x_k(\vp))
\ee
where
\be
\label{43}
V_k(x) = F_k(x,u_k) - F_{k-1}(x,u_k) + (u_k - u_{k-1}) \;
\frac{\prt}{\prt u_k} \; F_k(x,u_k) \; ,
\ee
with $u_k=u_k(x)$ and $k=1,2,\ldots$.

Our aim is to find a fixed point $y^*(\vp)$ of the map $\{ y_k(\vp)\}$, which, 
by construction, corresponds to the sought function $f(x)$. The fixed point for 
a flow is given by the zero velocity $v(y^*)=0$. For the approximation cascade, 
we have
\be
\label{44}
V_k(x) = 0 \; .
\ee
This condition is to be treated as an equation for control functions $u_k(x)$. 
However, the cascade velocity (43) contains two control functions, $u_k(x)$ and 
$u_{k-1}(x)$, hence one equation (44) cannot define them both. Thus, we either 
have to invoke some additional constraints or can use an approximate equation 
for finding control functions, for instance, by minimizing the absolute value 
of the cascade velocity
\be
\label{45}
|V_k(x)| \leq | F_k(x,u_k) - F_{k-1}(x,u_k)| + \left | (u_k - u_{k-1})
\frac{\prt}{\prt u_k}\; F_k(x,u_k) \right | \; .
\ee
There are three ways of formulating fixed-point conditions.

{\bf Fixed-point condition 1}. Find $u_1(x)$ from an additional condition, 
say, from the differential optimization condition (16) or from the difference 
optimization condition (20). Then, all other $u_k(x)$, with $k\geq 2$, are 
given by Eq. (44), that is, by the equation
\be
\label{46}
F_k(x,u_k) - F_{k-1}(x,u_k) +  (u_k - u_{k-1})
\frac{\prt}{\prt u_k}\; F_k(x,u_k) = 0 \; .
\ee
The disadvantage of this way is that control functions for $k=1$ and 
$k\geq 2$ are defined by different conditions. Also, if for some $k$ equation 
(46) has no solutions, then an ambiguity arises requiring some additional 
assumptions.

{\bf Fixed-point condition 2}. Minimizing the first term in the right-hand 
side of Eq. (45), one has
\be
\label{47}
F_k(x,u_k) - F_{k-1}(x,u_k) = 0 \; ,
\ee
which coincides with the difference condition (20). Then the cascade velocity 
(43) is
\be
\label{48}
V_k(x) = (u_k - u_{k-1})\frac{\prt}{\prt u_k} \; F_k(x,u_k) \; .
\ee
Since Eq. (47) gives $u_k(x)$ starting from $k=1$, the function $u_0(x)$ 
is left undefined. Hence, the velocity (48) is valid for $k\geq 2$. 
If Eq. (47) has no solution for some $k$, then one could find $u_k(x)$ by 
minimizing $|F_k-F_{k-1}|$. But in that case, the cascade velocity is not given 
by Eq. (48).

{\bf Fixed-point condition 3}. Minimize the second term in the right-hand 
side of Eq. (45), which yields
\be
\label{49}
(u_k - u_{k-1})\; \frac{\prt}{\prt u_k} \; F_k(x,u_k) = 0 \; .
\ee
This is to be understood as the equation
\be
\label{50}
\frac{\prt}{\prt u_k} \; F_k(x,u_k) = 0 \; , \qquad u_k\neq u_{k-1} \; ,
\ee
provided a solution for $u_k(x)$ exists, or as the equality
\be
\label{51}
u_k = u_{k-1} \; , \qquad \frac{\prt}{\prt u_k} \; F_k(x,u_k) \neq 0 \; ,
\ee
when the differential condition (50) does not possess a solution for $u_k(x)$. 
In all the cases under condition (49), the cascade velocity becomes
\be
\label{52}
V_k(x) = F_k(x,u_k) - F_{k-1}(x,u_k) \; ,
\ee
where $u_k=u_k(x)$ and $k\geq 1$. 

Comparing the possible fixed-point conditions (46), (47), and (49), we see that 
the latter is more general and provides a unique unambiguous way for defining 
control functions $u_k(x)$ for all $k\geq 1$. In all three cases, the function 
$u_0(x)$ is, generally, not defined, so is the term $f_0(x)=F_0(x,u_0(x))$. 
Therefore, the evolution integral (40) has to be considered starting with $k=2$. 
The effective time $\tau_k$ in Eq. (40) can be treated as another control function. 
By its definition, $\tau_k$ is the minimal time required for reaching a quasifixed 
point $f_k^*$ from an approximant $f_{k-1}$. In general, the minimal time should 
correspond to one step, which means that $k\tau_k$ should be close to one. 
Therefore, the effective approximation time can be evaluated as
\be
\label{53}
\tau_k = \frac{1}{k} \; .
\ee
In this way, an approximate fixed point $f_k^*(x)$, representing a $k$-th order 
{\it self-similar approximant} for the sought function $f(x)$, is completely 
defined by the evolution integral (40).

An important advantage of the self-similar approximation theory is the possibility 
of controlling the stability of the procedure, which can be done by invoking the 
ideas of dynamical theory. For this purpose, for a map $\{ y_k(\vp)\}$, one may 
define the {\it local multipliers}
\be
\label{54}
\mu_k(\vp) \equiv \frac{\prt}{\prt\vp} \; y_k(\vp) \; .
\ee
The multiplier at a quasifixed point $y_k^*(\vp)$ is given by
\be
\label{55}
\mu_k^*(\vp) \equiv \mu_k(y_k^*(\vp)) \; .
\ee
The images of these multipliers on the manifold $\Bbb{X}$ are obtained by means 
of the change of variables (23). The image of Eq. (54) reads
\be
\label{56}
m_k(x) \equiv \mu_k(F_0(x,u_k(x))) \; ,
\ee
and that of Eq. (55) is
\be
\label{57}
m_k^*(x) \equiv \mu_k(f_k^*(x)) \; .
\ee
A quasifixed point is stable provided that
\be
\label{58}
|\mu_k^*(\vp)| < 1 \; , \qquad | m_k^*(x) | < 1 \; .
\ee
It is useful to consider uniform stability characterized by the {\it maximal 
local multipliers}
\be
\label{59}
\mu_k^* \equiv \sup_\vp |\mu_k^*(\vp)| \; , \qquad 
m_k^* \equiv \sup_x | m_k^*(x) | \; .
\ee
Then a $k$-th order approximant is uniformly stable if
\be
\label{60}
\mu_k^* < 1 \; , \qquad  m_k^* < 1 \; .
\ee
Generally, because of the definition of the multipliers (55) to (57) through 
the same local multiplier (54) given on the phase space ${\cal A}$, the 
maximal multipliers (59) coincide, $\mu_k^*=m_k^*$, so that it is sufficient 
to consider one of them.

The stability analysis also makes it possible to answer the question "which 
of several admissible initial approximations should one prefer in calculating 
higher-order approximants?" The answer is straightforward: One has to prefer 
that initial approximation which garantees the best stability of the procedure 
[57].

In conclusion to this section, let us emphasize a principal aspect 
distinguishing our approach from different variants of the standard 
renormalization-group techniques [60--62]. In the latter, one tries to 
establish either an exact or an approximate relation between a function 
$f(x)$ and its value $f(\lbd x)$ for the scaled physical variable. Such a 
relation describes the motion with respect to the scaling parameter. Contrary 
to this, in self-similar approximation theory, we do not scale physical 
variables. But the group self-similarity describes an evolution in the phase 
space of approximants, with the approximation order playing the role of time.

\section{Method of Fractal Transforms}

One of the basic ideas in self-similar approximation theory is the 
introduction of control functions which govern the evolution of an approximation 
dynamical system to be close to a fixed point. As was mentioned earlier, control 
functions can be introduced at any part of calculational procedure. For instance, 
this can be done at the step of choosing an initial approximation, which results 
in the sequence of optimized approximants. But this can also be accomplished in 
the last part of calculations, after deriving a sequence of perturbative terms.

As is discussed in Section 1, employing the standard perturbation theory, one 
obtains approximations having the structure of asymptotic series. Such series 
are usually divergent and have no sense for finite values of expansion parameters. 
There exist the so-called resummation methods ascribing finite values to divergent 
series [63]. The most often used among such techniques are the Borel summation 
[63] and the construction of Pad\'e approximants [64], including the two-point 
[65] and multivalued [66,67] Pad\'e approximants. These techniques have many 
known limitations. Thus, to get a good accuracy, they require to invoke a number 
of perturbative terms which often are not available. And also, such techniques 
are, actually, numerical. 

In order to incorporate control functions into a given asymptotic series, 
one needs to resort to a transformation including some trial parameters. The 
transformation involved must decode the self-similarity property hidden in the 
given perturbative sequence. For power series, it looks natural to employ the 
power-law transformations [68--75]. Since power laws are typical of fractals
 [76,77] the power-law transformation can also be called the fractal 
transformation [75].

For a function $f(x)$, the {\it fractal transform} is 
\be
\label{61}
F(x,s) \equiv x^s f(x) \; ,
\ee
with a real $s$. The inverse transform is
\be
\label{62}
f(x) \equiv x^{-s} F(x,s) \; .
\ee
The fractal transform satisfies the scaling relation
$$
\frac{F(\lbd x,s)}{f(\lbd x)}  = \lbd^s\; \frac{F(x,s)}{f(x)}
$$
reminding us those typical of fractals.

Assume that for a finite function $f(x)$, the standard perturbation theory in 
powers of $x$ results in a set of approximations
\be
\label{63}
\vp_k(x) = \sum_{n=0}^k \vp_n x^{\gm_n} \; ,
\ee
where $\vp_0(x) =\vp_0 x^{\gm_0}\neq 0$ and $x\ra 0$. The powers are 
arranged in the ascending order
\be
\label{64}
\gm_n < \gm_{n+1} \qquad (n=0,1,2,\ldots,k-1) \; ;
\ee
the signs of $\gm_n$ being arbitrary. The structure of the series (63) is 
that of an expansion (6) over the asymptotic sequence $\ep_n(x) =x^{\gm_n}$.

Since it is always more convenient to work with dimensionless quantities, 
we assume that the variable $x$ is dimensionless. Define the dimensionless 
scale-invariant function
\be
\label{65}
g_k(x) \equiv \frac{\vp_k(x)}{\vp_0(x)} \; .
\ee
Employing the notation
$$
a_n \equiv \frac{\vp_n}{\vp_0}\; , \qquad 
\al_n \equiv \gm_n -\gm_0 \; ,
$$
we have
\be
\label{66}
g_k(x) = \sum_{n=0}^k a_n x^{\al_n} \; ,
\ee
where
\be
\label{67}
g_0(x) =a_0 = 1 \; \qquad \al_0 = 0 \; ,
\ee
and the powers are such that
\be
\label{68}
0 < \al_n < \al_{n+1} \qquad (n=1,2,\ldots,k-1) \; .
\ee

Introduce the {\it fractal transform}
\be
\label{69}
F_k(x,s) \equiv x^s g_k (x) 
\ee
and the inverse one
\be
\label{70}
g_k(x) = x^{-s} F_k(x,s) \; .
\ee
For the series (66), this gives
\be
\label{71}
F_k(x,s) = \sum_{n=0}^k a_n x^{s+\al_n} \; .
\ee
Then we apply the same procedure of self-similar approximation theory, 
described in the previous section, to the sequence $\{ F_k(x,s)\}$. The 
difference is that here the trial parameter $s$ is treated as a quasi-invariant, 
that is, it is kept fixed under the evolution of the map $\{ F_k(x,s)\}$ and 
it is transformed in a control function $s_k(x)$ aposteriori, the latter being 
defined from convergence and boundary conditions.

The reonomic constraint (23) now becomes
\be
\label{72}
F_0(x,s) =\vp \; , \qquad x=x(\vp,s) \; .
\ee
With the form (71), from where $F_0(x,s)=x^s$, this gives $x(\vp,s)=\vp^{1/s}$. 
The endomorphism (24) now is
\be
\label{73}
y_k(\vp,s) \equiv F_k(x(\vp,s), s) \; ,
\ee
which results in
\be
\label{74}
y_k(\vp,s) = \sum_{n=0}^k a_n \vp^{1+\al_n/s} \; .
\ee
For $s$ being a quasi-invariant, the cascade velocity (42) is
\be
\label{75}
v_k(\vp,s) = y_k(\vp,s) - y_{k-1}(\vp,s) \; .
\ee
From Eqs. (74) and (75), one gets
$$
v_k(\vp,s) = a_k \vp^{1+\al_k/s} \; .
$$
This is to be substituted in the evolution integral
\be
\label{76}
\int_{y_{k-1}^*}^{y_k^*} \frac{dy}{v_k(y,s)} = \tau_k \; ,
\ee
which is similar to the integral (39), and where $y_k^*=y_k^*(\vp,s)$ is a 
quasifixed point, with $y_0^*(\vp,s)\equiv \vp$ and $k\geq 1$. Note a slight 
difference between the integrals (39) and (76). In the present case, the 
evolution integral (76) is obtained by integrating the Lie equation (38) between 
two quasifixed points, $y^*_{k-1}$ and $y_k^*$. After changing variables according 
to the constraint (72), the integral (76) reduces to
\be
\label{77}
\int_{F_{k-1}^*}^{F_k^*} \frac{d\vp}{v_k(\vp,s)} = \tau_k \; ,
\ee
where $F_k^*=F_k^*(x,s)$, $k\geq 1$ and
\be
\label{78}
F_k^*(x,s) \equiv y_k^*(F_0(x,s),s) \; .
\ee
For $k=0$, since $y_0^*(\vp,s)=\vp$, one has $F_0^*(x,s)=x^s$.

Calculating the evolution integral (77), with the cascade velocity (75), 
results in the iterative equation
\be
\label{79}
\left ( F_k^* \right )^{\dlt_k} = 
\left ( F_{k-1}^*\right )^{\dlt_k} +A_k \; ,
\ee
in which
$$
A_k\equiv a_k\dlt_k\tau_k \; , \qquad \dlt_k\equiv - \frac{\al_k}{s} \; .
$$
If we recall that, to return to the original physical quantities, we have to 
accomplish the inverse fractal transform (70), then we define
\be
\label{80}
g_k(x,s) \equiv x^{-s} F_k^*(x,s) \; ,
\ee
with $g_0(x,s)=1$. For the transform (80), the iterative equation (79) reads
\be
\label{81}
g_k^{\dlt_k} = g_{k-1}^{\dlt_k} + A_k x^{\al_k} \; .
\ee

At this stage, we have to convert the trial parameter $s$ into a 
control function $s_k=s_k(x)$. Taking this into account, we come to a 
{\it self-similar approximant}
\be
\label{82}
g_k^*(x) \equiv g_k(x,s_k) \; .
\ee
Equation (81) for the approximant (82) can be written as
\be
\label{83}
g_k^*(x) =\left\{ \left [ g_{k-1}^*(x)\right ]^{\dlt_k} + 
A_k x^{\al_k} \right \}^{1/\dlt_k} \; ,
\ee
where $\dlt_k\equiv -\al_k/s_k$ and
\be
\label{84}
g_0^*(x) = 1 \; .
\ee
The solution to Eq. (83), with the initial condition (84), gives $g_k^*(x)$, 
which defines the self-similar approximant
\be
\label{85}
f_k^*(x) =\vp_0(x) g_k^*(x) 
\ee
for the sought function $f(x)$. The quantities $A_k$ and $\dlt_k$ are expressed
through control functions $\tau_k$ and $s_k$, which actually means that $A_k$
and $\dlt_k$ can be considered themselves as control functions. The latter are 
to be defined from additional conditions, such as convergence and boundary 
conditions.

\section{Self-Similar Root Approximants}

When the behaviour of the sought function is known in the asymptotic vicinity 
of two boundaries of the domain $\Bbb{X}$, the control parameters $A_k$ and 
$\dlt_k$ can be found from the related boundary conditions [71,73,74]. Here, 
we generalize this procedure to the case when the boundary asymptotic expansions 
contain arbitrary powers of $x$, including noninteger powers.

Suppose, for concreteness, that the variable $x$ is given on the real semiaxes 
$\Bbb{X}=\Bbb{R}_+\equiv [0,\infty)$. If this is not so, then it is always 
possible to resort to a change of variables reducing the domain of $x$ to 
$\Bbb{R}_+$. As earlier, we keep in mind that the variable $x$ as well as the 
sought function are normalized to dimensionless units, so that the considered 
function is presented in a scale-invariant form $g(x)$. Assume that the 
asymptotic behaviour of $g(x)$ in the vicinity of the left boundary,
\be
\label{86}
g(x) \simeq g_k(x) \qquad (x\ra 0) \; ,
\ee
is given by the asymptotic series
\be
g_k(x) = \sum_{n=0}^k a_n x^{\al_n} \; ,
\ee
where
\be
\label{88}
g_0(x) = a_0 = 1 \; , \qquad \al_0 = 0 \; ,
\ee
and the powers are arbitrary real numbers arranged in the ascending order
\be
\label{89}
0 < \al_n < \al_{n+1} \qquad (n=1,2,\ldots,k-1) \; .
\ee
And let us assume that the asymptotic behaviour of $g(x)$ at the right 
boundary is also known,
\be
\label{90}
g(x) \simeq G_k(x) \qquad (x\ra\infty) \; ,
\ee
being presented by the asymptotic series
\be
\label{91}
G_k(x) = \sum_{n=1}^k b_n x^{\bt_n} \; ,
\ee
in which
\be
\label{92}
b_1 \neq 0 \; , \qquad \bt_1 \neq 0 \; ,
\ee
and the powers are arranged in the descending order
\be
\label{93}
\bt_n > \bt_{n+1} \qquad (n=1,2,\ldots,k-1) \; .
\ee
Note that $\bt_n$ can be of any sign.

Iterating $k$ times Eq. (83), and using the notation
\be
\label{94}
n_p \equiv \frac{\dlt_{p+1}}{\dlt_p} \qquad (p=1,2,\ldots,k-1) \; , 
\qquad n_k \equiv \frac{1}{\dlt_k} \; ,
\ee
we obtain the {\it self-similar root approximant} 
\be
\label{95}
g_k^*(x) = \left (\ldots \left (\left ( 1 +A_1x^{\al_1}\right )^{n_1} +
A_2x^{\al_2}\right )^{n_2} + \ldots + A_k x^{\al_k}\right )^{n_k} \; ,
\ee
which can also be called a {\it nested root} or {\it superroot}. The control 
parameters $A_p$ and $n_p$, with $p=1,2,\ldots,k$, are to be defined from the 
asymptotic coincidence of expressions (95) and (91) at the right boundary, 
that is from the {\it asymptotic boundary condition}
\be
\label{96}
g_k^*(x) \simeq G_k(x) \qquad (x\ra\infty) \; .
\ee
In this way, the {\it crossover formula} (95) extrapolates the sought function
from the left boundary $x\ra 0$ to the whole interval $[0,\infty)$.

As is easy to observe, reexpanding Eq. (95) in small $x\ra 0$ does not 
reproduce the structure of $g_k(x)$ in the series (87). For this to hold, 
one, first, should require that $\al_p=p\al$, and even then the expansion 
coefficients would not coincide with $a_p$ for $p>1$. However, there is no 
need to demand that the asymptotic behaviour of $g^*(x)$ be identical with 
$g_k(x)$. Vice versa, such a restriction would essentially spoil the accuracy 
of the approximant for large $x$. It is important to bring to mind that the 
main aim of the self-similar approximation theory is to construct accurate 
expression {\it uniformly} approximating the sought function in the whole 
interval of the variable $x\in[0,\infty)$. The asymptotic expansion (87) at 
$x\ra 0$ are used only as constructing blocks. In general, it would be possible 
to modify the form of the self-similar approximant so that it could exactly 
reproduce $g_k(x)$ at $x\ra 0$. But this, from the point of view of 
an accurate uniform extrapolation, is neither necessary nor correct.

At the same time, by the asymptotic condition (96), the control parameters 
$A_p$ and $n_p$ are defined so that the asymptotic expansion of $g_k^*(x)$ 
at $x\ra\infty$ exactly coincide with the series (91). Such a construction, 
as is evident, can be reverted in the following sense. One could derive 
self-similar root approximants starting from the right asymptotic expansion (91)
and fitting the corresponding control parameters so that the expansion of the 
derived root approximants at $x\ra 0$ be coinciding with the series (87). 
Nevertheless, the construction of crossover formulas from the left to right 
seems more preferable because of the following. The region of validity for 
the expansion (87) is $|x|\ll 1$ and that for the expansion (91) is $|x|\gg 1$, 
that is, the region of validity of the right expansion is essentially larger 
than that of the left expansion. This conclusion is confirmed by a number of 
particular cases demonstrating that the radius of convergence of the series 
(87) is usually zero, while that of the series (91) is finite.

In order to define the control parameters $A_p$ and $n_p$ from the asymptotic
condition (96), one has to know how to present an asymptotic form of the root
approximant (95) at $x\ra\infty$. To the first glance, the procedure of
obtaining an asymptotic, as $x\ra\infty$, expression from the superroot (95)
looks ambiguous, since the powers $n_p$ are yet not known, hence it is not 
clear how to classify larger and smaller terms. In particular cases, really, 
there can be several ways, depending on the relation between the values of 
$\al_p$ and $n_p$, of classifying asymptotic terms of the superroot (95) at 
$x\ra\infty$. This ambiguity can be overcome by requiring the uniqueness and 
generality of the procedure.

\vskip 2mm

{\bf Definition}. The self-similar root approximant (95) is called to be 
uniquely defined by the asymptotic condition (96) if and only if all control 
parameters can be uniquely determined from a general rule, whose form is 
invariant with respect to the values of $\al_p$ and which is valid for 
arbitrary $p=1,2,\ldots$.

\vskip 2mm

{\bf Theorem}. The self-similar root approximant (95) is uniquely defined 
by the asymptotic condition (96) if and only if the powers $n_p$ are given 
by the equations 
$$
\al_p n_p -\al_{p+1}  = const \; ,
$$
\be
\label{97}
\al_p n_p = \al_{p+1} - \bt_{k-p} + \bt_{k-p+1} \; ,
\ee
$$
\al_kn_k =\bt_1 \qquad (p=1,2,\ldots,k-1) \; .
$$

\vskip 1mm

{\bf Proof}. When $x\ra\infty$, it is convenient to introduce the small 
parameter $\ep\equiv x^{-1}$. In terms of the latter, the superroot (95) 
can be identically presented as
$$
g_k^*(x) = A_k^{n_k} x^{\al_k n_k} \left ( 
1 + B_k\ep^{\al_k-\al_{k-1}n_{k-1}}
\left ( 1 + B_{k-1}\ep^{\al_{k-1} -\al_{k-2} n_{k-2}}\left ( 1 + \ldots 
\right. \right. \right.
$$
\be
\label{98}
\left. \left. \left. + B_2 \ep^{\al_2 -\al_1 n_1}\left ( 
1 + B_1 \ep^{\al_1} \right )^{n_1} \right )^{n_2} \ldots 
\right )^{n_{k-1}} \right )^{n_k} \; ,
\ee
where
$$
B_1 \equiv \frac{1}{A_1} \; , \qquad 
B_{p+1} \equiv \frac{A_p^{n_p}}{A_{p+1}} \qquad (p=1,2,\ldots,k-1) \; .
$$
To prove {\it sufficiency}, let us assume that Eqs. (97) hold true. Then,
$$
\al_{p+1} - \al_p n_p = \bt_{k-p} - \bt_{k-p+1} \; .
$$
Due to the descending order (93) of $\bt_n$, one has
$$
\al_{p+1} - \al_p n_p = const > 0 \; .
$$
This unambiguously defines the classification of powers of $\ep$ in the 
form (98). Expanding Eq. (98) in powers of $\ep$, we observe that the first 
$k$ terms of the expansion coincide with all $k$ terms of the series (91), 
that is the asymptotic condition (96) is satisfied. To prove {\it necessity}, 
we assume that the superroot (95) is uniquely defined by the asymptotic condition 
(96). This implies, according to the above definition, that there exists a unique 
general expansion of the form (98) in powers of $\ep$, which is invariant with 
respect to $\al_p$ and $p$. As is evident from expression (98), such a unique 
general expansion is possible if and only if $\al_{p+1} - \al_p n_p = const >0$.
Then Eq. (98) can be unambiguously expanded in powers of $\ep$. This expansion 
is to be compared with the series (91) that can be written as
$$
G_k(x) = b_1x^{\bt_1} \left ( 1 + \frac{b_2}{b_1}\; \ep^{\bt_1-\bt_2} 
\left ( 1 + \frac{b_3}{b_2} \; \ep^{\bt_2-\bt_3}\left ( 1 + \ldots +
\frac{b_k}{b_{k-1}}\; \ep^{\bt_{k-1}-\bt_k}\right )\right ) 
\ldots \right )  \; .
$$
Comparing the first $k$ terms of the expansion of Eq. (98) with the series 
(91), we obtain Eqs. (97).

This theorem makes it possible to apply the same general rules for 
constructing various crossover formulas. Let us stress that the theorem is 
a new result that has not yet been published.

Till now, we have considered the situation when the order $k$ of the left 
expansion (87) coincides with that of the right expansion (91). How could 
we proceed if these orders were different?

If the number of available terms from the left is less than that from the 
right, this is not as important, provided we know the law prescribing the 
values of $\al_n$, which is usually known or can be easily guessed.
This is because we, actually, do not need to have all coefficients $a_n$, 
which are incorporated in the control parameters $A_n$, and these are 
determined through the coefficients $b_n$ and powers $\bt_n$ of the right 
expansion (91). Therefore, when only $\al_n$ are available, nevertheless, 
we may add to the left expansion the required number of terms up to the 
order of the right expansion.

When the number of terms in the right expansion is less than that of 
the left expansion, the situation again is not dangerous. Say, the left 
expansion $g_k(x)$ is of order $k$, while the right expansion $G_m(x)$ is 
of order $m<k$. In that case, we iterate Eq. (83) till $g_{k-m+1}^*(x)$ 
and at the next step, we set $g_{k-m}(x)$. Iterating in this way Eq. (83) 
$m$ times, we come to the self-similar root approximant
$$
g_{km}^*(x) = \left ( \ldots \left (\left ( \left [ 
g_{k-m}(x)\right ]^{1/n_{k-m+1}} + A_{k-m+1} x^{\al_{k-m+1}} 
\right )^{n_{k-m+1}} + \right. \right. 
$$
\be
\label{99}
\left. \left. + A_{k-m+2} x^{\al_{k-m+2}} \right )^{n_{k-m+2}} + 
\ldots + A_k x^{\al_k} \right )^{n_k} \; ,
\ee
where the notation (94) is used. The superroot (99) for the case $m=k$ 
returns to the form (95), since $g_0(x)=1$. The control parameters $A_p$ 
and $n_p$, with $p=k-m+1,\; k-m+2,\ldots,k$, are defined by the 
{\it asymptotic boundary condition} 
\be
\label{100}
g_{km}^*(x) \simeq G_m(x) \qquad (x\ra\infty) \; .
\ee
Consequently, the self-similar root approximants can always be constructed, 
even when the number of terms in the left and right asymptotic expansions 
are not equal to each other.

\section{Self-Similar Exponential Approximants}

A different strategy is to be pursued when only a single-side asymptotic 
expansion, say at $x\ra0$, is available. Then there are no boundary conditions 
determining control parameters. The latter are to be specified in a different 
way appealing to convergence properties. A particular choice of control parameters 
results in a nice structure of nested exponentials [70,72,75]. Here we present a 
more refined derivation of the exponential approximants and suggest some novel 
ways of constructing the cost functionals defining the effective approximation 
time, that is, the control time.

Looking at the fractal transform (71), it is easy to notice that the 
convergence properties of the sequence $\{ F_k\}$ improve if $|x|^s\ra 0$. 
The latter can be realized when $x\ra 0$, i.e. in the same situation as for 
the asymptotic expansion (66). But there are in addition two other possibilities 
for $|x|^s$ to tend to zero, when
\begin{eqnarray}
\label{101}
s\ra \left\{ \begin{array}{cc}
+\infty\; , & |x| < 1 \\
-\infty\; , & |x| > 1 \; . 
\end{array} \right.
\end{eqnarray}
By taking these limits, we may extend the region of applicability of the 
function presented by the asymptotic expansion (66), valid only at $x\ra 0$, 
to the regions $[0,1)$ and $(1,\infty)$.

To derive a self-similar approximant for the case (101), by employing the 
method of fractal transforms, we need to obtain a kind of an iterative equation, 
similar to Eq. (83). For this purpose, let us introduce a set of functions
$$
\psi_k(z_n) \equiv 1 + z_n \qquad (n=1,2,\ldots,k) \; ,
$$
\be
\label{102}
z_n = z_n(x) \; , \qquad \psi_k(z_{k+1}) \equiv 1 \; ,
\ee
being iteratively connected with each other by means of the relation
\be
\label{103}
z_n(x) = \frac{a_n}{a_{n-1}}\; x^{\al_n -\al_{n-1}}\; \psi_k (z_{n+1}) \; .
\ee
Then the series (66) can be identically presented as
\be
\label{104}
g_k(x) = \psi_k(z_1) \; .
\ee
The self-similar renormalization of $\psi_k(z_n)$, accomplished by means of 
the method of fractal transforms, is
\be
\label{105}
\psi_k^*(z_n,s) =\left ( 1 -\; \frac{1}{s}\; \tau_n z_n \right )^{-s} \; .
\ee
Realizing the $k$-step renormalization for the iterative relations (102) and
(103), we find
\be
\label{106}
g_k(x,s) \equiv \psi_k^*(z_n^*,s) \; ,
\ee
where $z_n^*=z_n^*(x,s)$, with $n=1,2,\ldots,k$, and
\be
\label{107}
z_n^*(x,s) =\frac{a_n}{a_{n-1}}\; x^{\al_n-\al_{n-1}} \; 
\psi_k^*(z_{n+1}^*,s) \; .
\ee
According to the last identity in Eqs. (102),
$$
\psi_k^*(z_{k+1}^*,s) =  1 \; .
$$

Irrespectively to what limit, either $s\ra +\infty$ or $s\ra -\infty$, is 
taken in the form (105), one gets the same result
$$
\lim_{s\ra\pm\infty}\left ( 1 -\; \frac{1}{s}\;\tau z\right )^{-s} =
\exp(\tau z) \; .
$$
Therefore, in what follows, we may write $|s|\ra\infty$, keeping in mind any 
of the limits $s\ra\pm\infty$. Let us define
\be
\label{108}
g_k^*(x) = \lim_{|s|\ra\infty} g_k(x,s) \; .
\ee
And introduce the notation
\be
c_n\equiv \frac{a_n}{a_{n-1}}\; \tau_n \; , \qquad
\nu_n \equiv \al_n -\al_{n-1} \; .
\ee
Taking the limit (108) in the iterative relations (106) and (107), we come
to the {\it self-similar exponential approximant}
\be
\label{110}
g_k^*(x) =\exp\left ( c_1 x^{\nu_1}\exp\left ( c_2 x^{\nu_2}\ldots
\exp\left ( c_k x^{\nu_k}\right ) \right ) \ldots \right ) \; ,
\ee
for short, called {\it superexponential}.

Expression (110) contains the coefficients $c_n$ that, as is seen from the
notation (109), are proportional to the control time $\tau_n$, which is not 
yet defined. The simplest way would be to set $\tau_n=1/n$, as in Eq. (53). 
It could also be possible to find $\tau_n$ from a fixed-point condition. 
However, the most general and refined way is to determine the control time 
by minimizing a cost functional [75]. In optimal control theory, one 
constructs cost functionals by formulating the desired properties of the 
system. For our case, the procedure can be as follows. If it is recalled 
that the control time $\tau_n$ describes the minimal time necessary for 
reaching a fixed point at the $n$-th step of the calculational procedure, 
then $n\tau_n$ approximately corresponds to the total time $\tau^*$ required 
for reaching the fixed point. When $n\tau_n\sim\tau^*$, this implies that 
$\tau_n\sim\tau^*/n$. The time of reaching a fixed point depends on how far 
this point is. The shorter is the distance from the point, the faster is 
the way to it. The distance passed at the $n$-th step can be evaluated
as $v_n^*\tau_n$, with $v_n^*$ being a characteristic velocity at this step.
In this manner, we need to find a minimal time $\tau_n$ that is close to 
$\tau^*/n$ and which corresponds to the fastest passage of the distance 
$v_n^*\tau_n$. These requirements suggest to construct the {\it fastest 
passage cost functional}
\be
\label{111}
{\cal F}_\tau = \frac{1}{2} \sum_n \left [ \left ( \tau_n -\;
\frac{\tau^*}{n}\right )^2 + 
\zeta\left (v_n^*\tau_n\right )^2 \right ] \; ,
\ee
in which the parameter $\zeta\geq 0$ is included for generality. The value 
of $\zeta$ can be chosen if some additional information on the system is 
available. In the absence of such an additional information, we set $\zeta=1$.

Defining the characteristic velocity $v_n^*$, it is natural to associate it
with a cascade velocity $v_n(x)$ taken at the most dangerous value of $x$, 
where convergence is the worst and, respectively, the deviation $v_n^*\tau_n$ 
should be the largest. Thinking back to the fractal transform (71), we know 
that the sequence $\{ F_k\}$ converges, under condition (101), if either 
$|x|<1$ or $|x|>1$. This means that the dangerous point is $|x|=1$. Therefore, 
we define
\be
\label{112}
v_n^* = v_n(x) \qquad (|x|=1) \; .
\ee
With the cascade velocity
\be
\label{113}
v_n(x) = g_n(x) - g_{n-1}(x) = a_n x^{\al_n} \; ,
\ee
we have
\be
\label{114}
\left ( v_n^*\right )^2 = a_n^2 \; .
\ee
So that the fastest passage cost functional (111) becomes
\be
\label{115}
{\cal F}_\tau = \frac{1}{2} \sum_n \left [ \left ( \tau_n -\; 
\frac{\tau^*}{n} \right )^2 + \zeta a_n^2 \tau_n^2\right ] \; .
\ee

The control time $\tau_n$ is given by the minimization of the cost 
functional (115), i.e. from the conditions
\be
\label{116}
\frac{\dlt{\cal F}_\tau}{\dlt\tau_n} = 0 \; ,\qquad
\frac{\dlt^2{\cal F}_\tau}{\dlt\tau_n^2} > 0 \; .
\ee
The extremum condition leads to
\be
\label{117}
\tau_n = \frac{\tau^*}{n(1+\zeta a_n^2)} \; .
\ee
The found extremum is a minimum, since
$$
\frac{\dlt^2{\cal F}_\tau}{\dlt\tau_n^2} = 1 +\zeta a_n^2 > 0 \; .
$$

What has been yet left undefined is the effective total time $\tau^*$, 
which can be derived from the following reasoning. If the sought fixed 
point is reached in one step, this implies that $\tau_1=1$. Applying this 
condition to formula (117) yields
\be
\label{118}
\tau^* = 1 +\zeta a_1^2 \qquad (\tau_1=1) \; .
\ee
Then the {\it control time} is
\be
\label{119}
\tau_n = \frac{1+\zeta a_1^2}{n(1+\zeta a_n^2)} \; .
\ee
The parameters $c_n$, defined in Eq. (109), are proportional to the 
control time $\tau_n$, because of which they can be called the control 
parameters or, simply, controllers. With the control time (119), the 
{\it controllers} are
\be
\label{120}
c_n = \frac{a_n(1+\zeta a_1^2)}{na_{n-1}(1+\zeta a_n^2)} \; .
\ee
In this way, the superexponential (110) is completely defined.

If the function $g(x)$ was introduced as a scale-invariant form of the 
sought function $f(x)$, then the self-similar exponential approximant for 
the latter is $f_k^*(x) =\vp_0(x)g_k^*(x)$. Recall that the function $g(x)$ 
has been assumed to be finite on the manifold $\Bbb{X}$. In the case of a 
function $g(x)$ divergent at some point $x_0\in\Bbb{X}$, one should consider 
its inverse $g^{-1}(x)$, provided this is a finite everywhere on $\Bbb{X}$. 
In the example of deriving the control time (119) from the cost functional 
(115), it is supposed that the function $f(x)$ is sign definite so that the 
function $g(x)$ is nonnegative. When it is known that $f(x)$ changes its 
sign, this information has to be encompassed in the procedure. This can be 
done, for example, by factoring $f(x)=\vp(x)g(x)$, with $g(x)$ being positive. 
Another possibility could be to incorporate information on the points of the 
sign change into the constructed cost functional. For describing oscillating 
functions, it could be conceivable to deal with complex control times.

\section{Examples}

Self-similar approximation theory has been applied to various physics 
problems. These applications can be found in the cited references 
[35--57,68--75]. Among recent works, we may mention the usage of this 
approach to barrier crossing processes [78,79], critical phenomena [80], 
and to the rupture of mechanical systems [81]. In the present section, 
we give several examples which illustrate some new possibilities of the 
approach.

\vskip 2mm

{\bf A. Amplitude of Elastic Scattering}

\vskip 1mm

This example is interesting by demonstrating the use of superroots when 
the number of terms in the left asymptotic expansion is much larger than 
that in the right expansion.

Consider the scattering of two particles of masses $m_1$ and $m_2$, 
with momenta $p_1$ and $p_2$ before collision and $p_1'$ and $p_2'$ 
after it. The four-momenta are normalized on the mass shell so that 
$p_i^2=m_i^2$. The scattering amplitudes  are usually presented as 
functions of the Mandelstam variables [82] which are
$$
s \equiv (p_1 + p_2)^2 =  (p_1' + p_2')^2 \; , \qquad
t \equiv (p_1 - p_1')^2 =  (p_2 - p_2')^2 \; ,
$$
$$
u \equiv (p_1 - p_2')^2 =  (p_2 - p_1')^2 \; , \qquad
s + t + u = 2(m_1^2 + m_2^2 ) \; .
$$
The amplitude of elastic scattering can be expressed, be means of 
perturbation theory, as an asymptotic expansion in powers of the coupling 
parameter $g$,
\be
\label{121}
T(g,s,t) \simeq g + 
\sum_{n=2}^\infty T_n(s,t) g^n \qquad (g\ra 0) \; ,
\ee
where $T_n(s,t)\sim\exp(ns)$ as $s\ra\infty$. It is known that for any 
$g$ there exists the Froissart upper bound given by the inequality 
$|T(g,s,t)|\leq |A(s,t)|$, where a particular form of $A(s,t)$ depends 
on whether the considered theory is local [82] or nonlocal [83]. Since 
the Froissart upper bound is valid for any $g$, including $g\ra\infty$, 
let us assume that
\be
\label{122}
T(g,s,t) \simeq A(s,t) \qquad (g\ra\infty) \; .
\ee
Our aim is to construct a crossover formula between the left and 
right expansions (121) and (122), respectively.

Following the general scheme, we, first, have to introduce the 
scale-invariant function
\be
\label{123}
f(g) \equiv \frac{1}{g}\; T(g,s,t) \; ,
\ee
in which, for brevity, we do not write explicitly other variables, 
except $g$. Denoting for the function (123) the weak-coupling,
\be
\label{124}
f(g) \simeq \vp_k(g) \qquad (g\ra 0) \; ,
\ee
and the strong-coupling,
\be
\label{125}
f(g) \simeq F_1(g) \qquad (g\ra \infty) \; ,
\ee
asymptotic expansions, from Eqs. (121) and (122), we have
\be
\label{126}
\vp_k(g) = 1 +a_1 g + a_2 g^2 + \ldots + a_k g^k \; , \qquad
F_1(g) = bg^{-1} \; ,
\ee
where $a_n=T_{n+1}(s,t)$ and $b=A(s,t)$.

The crossover formula for the scattering amplitude
\be
\label{127}
T_k^*(g) = g\vp^*_{k1}(g)
\ee
is obtained by constructing the self-similar root approximant (99) 
for
\be
\label{128}
\vp_{k1}^*(g) =\left ( \left [ \vp_{k-1}(g)\right ]^{1/n_k} +
A_k g^k \right )^{n_k} \; ,
\ee
where the control parameters $n_k$ and $A_k$ are defined by the 
asymptotic boundary condition (100), which gives
\be
\label{129}
n_k = -\;\frac{1}{k} \; , \qquad A_k = \frac{1}{b^k} \; .
\ee
Thus, a self-similar root approximant for the scattering amplitude is
\be
\label{130}
T_k^*(g) =\frac{b\vp_{k-1}(g)g}{[b^k +\vp_{k-1}^k(g)g^k]^{1/k}}\; .
\ee
Note that the expansion $\vp_{k-1}(g)$ here can also be converted to 
$\vp_{k-1}^*(g)$ given by the superexponential (110).

\vskip 2mm

{\bf B. Summation of Numerical Series}

\vskip 1mm

Poorly convergent or divergent numerical series can be summed by means
of the superexponentials in the following way. Let us consider a series
$S_\infty=\sum_{n=0}^\infty a_n$, whose particular sums are
\be
\label{131}
S_k = \sum_{n=0}^k a_n \; .
\ee
Introduce the function
\be
\label{132}
S_k(x) \equiv \sum_{n=0}^k a_n x^n \; ,
\ee
for which $S_k(1)=S_k$. Construct the self-similar exponential approximant
$S_k^*(x)$, according to formula (110). Setting $x=1$ in $S_k^*(x)$, we 
get $S_k^*=S_k^*(1)$, which is the sought self-similar approximant 
\be
\label{133}
S_k^* = a_0\exp\left (c_1\exp\left (c_2\ldots \exp(c_k)\right )
\ldots \right ) \; ,
\ee
with the controllers $c_n$ given by Eq. (120), where we set $\zeta=1$.
If the exact value of $S_\infty$ is known, one can compare the accuracy
of the particular sums (131), characterized by the percentage error
\be
\label{134}
\ep_k \equiv \left ( \frac{S_k}{S_\infty} \; - 1\right ) \cdot
100\% \; ,
\ee
with the accuracy of the self-similar approximants (133), described by 
the error 
\be
\label{135}
\ep_k^* \equiv \left ( \frac{S_k^*}{S_\infty} \; - 1\right ) \cdot
100\% \; .
\ee

As an illustration, let us consider the sum (131), with the coefficients
$$
a_n = \frac{(-1)^n}{2n+1} \; .
$$
The sequence $\{ S_k\}$ converges to $S_\infty =\pi/4$. This convergence 
is rather slow, for example, the percentage errors (134) for the first 
five terms are 
$$
-15\% \; , \quad 10\%\; , \quad -7.8\% \; , \quad
6.3\% \; , \quad -5.3\% \; ,
$$
while the superexponential (133) gives the errors (135) for the first 
five approximants as 
$$
-8.8\% \; , \quad -3.6\% \; , \quad  -1.9\% \; , \quad
-1.6\% \; , \quad -1.6\% \; ,
$$ 
demonstrating a much faster convergence.

\vskip 2mm

{\bf C. Multiloop Feynman Integrals}

\vskip 1mm

Employing the Feynman diagram techniques in quantum field theory or 
quantum statistical mechanics, one confronts with the so-called multiloop 
integrals. These can be calculated by means of perturbation theory [84,85] 
resulting in asymptotic series. The latter can be summed with the help of 
the superexponentials.

Let us start the illustration with a simple one-loop integral
\be
\label{136}
I(a,D) \equiv \frac{1}{(2\pi)^D}\;
\int \frac{d^Dp}{(1+{\bf p}^2)^a} \; ,
\ee
where $a$ is a positive parameter, $D$ is space dimensionality. The exact
value of the integral (136) is known to be
\be
\label{137}
I(a,D) = \frac{\Gamma(a-D/2)}{(4\pi)^{D/2}\Gamma(a)}\; ,
\ee
where $\Gamma(\cdot)$ is a gamma-function. A perturbative procedure for 
Eq. (136) can be defined [84,85] by introducing
\be
\label{138}
I(a,D,\ep) \equiv \frac{1}{(2\pi)^D} \int 
\frac{d^D p}{(1+\ep{\bf p}^2)^{a/\ep}} 
\ee
and expanding the integrand in powers of $\ep$, which, after the 
integration term by term, results in a series
\be
\label{139}
I(a,D,\ep) = \sum_n \vp_n \ep^n \; .
\ee
Since, as follows from Eqs. (136) and (138), $I(a,D,1)= I(a,D)$, the 
answer is obtained by setting $\ep=1$.

Accomplishing a partial self-similar exponentization of the series 
(139), one gets
\be
\label{140}
I^*(a,D,\ep) = \frac{1}{(4\pi a)^{D/2}}\; 
\exp\left\{ \frac{D(D+2)}{8a}\; \ep g(\ep) \right \} \; ,
\ee
with
\be
\label{141}
g(\ep) \equiv \sum_n a_n \ep^n \; ,
\ee
the coefficients $a_n$ being
$$
a_0 = 1 \; , \qquad a_1 =\frac{D+1}{6a} \; , \qquad 
a_2 =\frac{D(D+2)}{24a^2} \; ,
$$
$$
a_3 = \frac{(D+1)(3D^2+6D-4)}{240 a^3} \; , \qquad
a_4 = \frac{D(D+2)(D^2+2D-2)}{240 a^4} \; , \qquad \ldots
$$
The partial sums of Eq. (141), after setting $\ep=1$, become
\be
\label{142}
g_k=\sum_{n=0}^k a_n \; .
\ee
The corresponding superexponentials are
\be
\label{143}
g_k^* =\exp\left ( c_1\exp\left ( c_2 \ldots \exp(c_k)
\right ) \right )  \; .
\ee
Finally, for the integral (136), we find the self-similar approximants
\be
\label{144}
I_k^*(a,D) = \frac{1}{(4\pi a)^{D/2}}\; \exp\left\{ 
\frac{D(D+2)}{8a}\; g_{k-1}^*\right \}  \; ,
\ee
with $g_0^*=1$.

Consider the case of $a=1$ and $D=1$, when $I(1,1)=1/2$. The perturbation
series (139) take the form
\be
\label{145}
I(1,1,\ep) \simeq \frac{1}{\sqrt{4\pi}} \left ( 1 +
\frac{3}{8}\;\ep + \frac{25}{128}\;\ep^2 + \frac{105}{1024}\; \ep^3 +
\frac{1659}{32768}\; \ep^4\right ) \; .
\ee
The coefficients $a_n$ in the sum (142) are 
$$
a_0=1 \; , \quad a_1=\frac{1}{3}\; ,  \quad a_2=\frac{1}{8}\; ,
\quad a_3=\frac{1}{24}\; , \quad  a_4=\frac{1}{80} \; .
$$ 
For the self-similar approximant (144),
we have
\be
\label{146}
I_k^*(1,1) =\frac{1}{\sqrt{4\pi}}\; 
\exp\left ( \frac{3}{8}\; g_{k-1}^*\right ) \; .
\ee
The errors of the perturbative expression (145) at $\ep=1$ are
$$
-44\%\; , \quad -22\%\; , \quad -11\%\; , \quad -5.6\%\; , \quad
-2.8\% \; ,
$$
which is to be compared with the errors of the self-similar approximants 
(146),
$$
-7\%\; , \quad -4.8\%\; , \quad -0.77\%\; , \quad -0.14\%\; , \quad
-0.08\% \; ,
$$
which are an order smaller.

For the case $a=2,\; D=3$, one has $I(2,3)=1/8\pi$. The perturbative 
expression (139) reads
\be
\label{147}
I(2,3,\ep) \simeq \frac{1}{(8\pi)^{3/2}} \left ( 1 +\frac{15}{16}\;\ep +
\frac{385}{512}\;\ep^2 + \frac{4725}{8192}\;\ep^3 +
\frac{228459}{524288}\;\ep^4\right ) \; .
\ee
The coefficients $a_n$ from Eq. (142) are
$$
a_0=1 \; , \quad a_1 =\frac{1}{3}\; , \quad a_2=\frac{5}{32} \; \quad
a_3 =\frac{41}{480}\; , \quad a_4 =\frac{13}{256} \; .
$$
The self-similar approximant (144) becomes
\be
\label{148}
I_k^*(2,3) =\frac{1}{(8\pi)^{3/2}}\; 
\exp \left ( \frac{15}{16}\; g_{k-1}^*\right ) \; .
\ee
The direct expansion (147) yields the errors
$$
-80\%\; , \quad -61\%\; , \quad -46\%\; , \quad -35\%\; , \quad
-26\% \; ,
$$
while those of the self-similar approximants (148) are lower:
$$
-30\%\; , \quad -26\%\; , \quad -16\%\; , \quad -13\%\; , \quad
-12\% \; .
$$

Now, let us turn to a $D$-dimensional three-loop Feynman integral
\be
\label{149}
J(D) \equiv \frac{1}{(2\pi)^{3D}} \int
\frac{d^Dp_1\; d^Dp_2\; d^Dp_3}{(1+{\bf p}_1^2)(1+{\bf p}_2^2)(1+{\bf p}_3^2)
[1+({\bf p}_1+{\bf p}_2+{\bf p}_3)^2]} \; .
\ee
Following the same procedure as in the calculation of the previous Feynman 
integrals, one defines $J(D,\ep)$ and then set $\ep=1$. For concreteness, let 
us take $D=2$. Then the self-similar approximants for integral (149) are 
defined as
\be
\label{150}
J_k^*(2) = \frac{1}{256\pi^3} \; \exp\left ( \frac{9}{4}\;
g_{k-1}^*\right ) \; ,
\ee
with $g_k^*$ having the form (143), which is obtained from $g_k$ of 
Eq. (142), where the coefficient $a_n$ are
$$
a_0=1 \; , \quad a_1 =\frac{7}{24}\; , \quad a_2=\frac{13}{144} \; \quad
a_3 =\frac{59}{768}\; , \quad a_4 =\frac{373}{3840} \; ,
$$
$$
a_5 =\frac{2324}{18432}\; , \qquad a_6=\frac{15243}{86016} \; \qquad
a_7 =\frac{150379}{393216}\; .
$$
The accuracy of the approximants (150) again is much better than that of 
simple perturbative expressions. The first seven approximants demonstrate a 
fast increase of accuracy. The related errors, calculated by comparing the 
self-similar form (150) with the numerical value $J(2)=0.00424027$, are
$$
-46\%\; , \quad -40\%\; , \quad -29\%\; , \quad -24\%\; , \quad
-21\% \; , \quad -20\% \; , \quad -19\% \; .
$$
This demonstrates a monotonic convergence, while the standard perturbation
theory in powers of $\ep$ wildly diverges.

Thus, the self-similar approximants provide rather good approximations
even for very bad, fastly divergent series derived by means of standard
perturbation theory.

\vskip 2mm

In conclusion, we may mention that the self-similar approximation 
theory has been successfully applied not only to a number of physical 
problems [35--57,68--75,78--81] but also to other complex systems, such as 
financial markets [86--90]. Time series, related to financial, economic, 
biological, and social systems, are known to possess special fractal properties 
[91--95]. This is why these series can be naturally described by self-similar 
approximants, especially by those that explicitly display their self-similar 
structure as in self-similar roots and self-similar exponentials.

\end{document}